\documentclass{moriond}

\def\Journal#1#2#3#4{{#1} {\bf #2}, #3 (#4)}

\def\NIMA{{\em Nucl. Instrum. Methods} A}

\def\PRL{\em Phys. Rev. Lett.}
\def\PRD{{\em Phys. Rev.} D}

\def\be{\begin{equation}}
\def\ee{\end{equation}}
\def\bea{\begin{eqnarray}}
\def\eea{\end{eqnarray}}

\usepackage{lineno}
\usepackage{amssymb}
\usepackage[superscript]{cite} 
\usepackage[sort&compress,numbers]{natbib}
\begin{document}
\vspace*{4cm}
\title{New Measurement of Muon Neutrino Disappearance from the IceCube Experiment}

\author{ Shiqi Yu, for the IceCube Collaboration \footnote{\protect\url{http://icecube.wisc.edu}}}

\address{Dept. of Physics and Astronomy, Michigan State University, East Lansing, MI 48824, USA}

\maketitle\abstracts{
The IceCube Neutrino Observatory is a Cherenkov detector located at the South Pole. Its main component consists of an in-ice array of optical modules instrumenting one cubic kilometer of deep Glacial ice. The DeepCore sub-detector is a denser in-fill array with a lower energy threshold, allowing us to study atmospheric neutrinos oscillations with energy below 100 GeV arriving through the Earth. We present preliminary results of an atmospheric muon neutrino disappearance analysis using data from 2012 to 2021 and employing convolutional neural networks (CNNs) for precise and fast event reconstructions.}

%The IceCube Neutrino Observatory is a Cherenkov detector located deep under the South Pole ice. The main IceCube array instrumenting over one cubic kilometer of antarctic ice can detect high-energy astrophysical neutrinos. The DeepCore sub-detector has a denser configuration and can see down to GeV-scale neutrinos, allowing us to study $\nu_\mu$ disappearance using atmospheric neutrinos with energy below 100 GeV arriving through the Earth. In this analysis, convolutional neural networks are employed for precise and fast reconstructions. The result of using the data taken between 2012 and 2021 is discussed and compared to the existing measurements from the other experiments.}

\section{Introduction}
The phenomenon of neutrino oscillation has been observed in many experiments~\cite{icecube_prd,icecube_prl,nova_paper,t2k_paper,minos_paper,dayabay_paper}. Neutrinos that are produced in weak interactions as flavor eigenstates ($\nu_{e,\mu,\tau}$) start to oscillate between flavors due to their non-trivial mixing with neutrino mass eigenstates ($\nu_{1,2,3}$). The mixing is described by a unitary matrix that - assuming Dirac neutrinos - can be parameterized by three mixing angles ($\theta_{12}$, $\theta_{13}$, and $\theta_{23}$) and one CP-violating phase $\delta_{CP}$. The frequency of the flavor oscillation is determined by the mass splitting $\Delta m_{ij} \equiv m^2_{i}$-$m^2_j$ between the three neutrino masses $m_i$. Some parameters, such as neutrino mass splitting ($\Delta m^2_{21} \approx 7.4\times 10^{-5}\rm{eV}^2/c^4$) or the mixing angles $\theta_{13}\approx9^{\circ}$ and $\theta_{12}\approx34^{\circ}$~\cite{pdg} are well constrained by existing measurements, while the measurements of $\Delta m^2_{32}$ and $\theta_{23}$ have large uncertainties. These parameters can be constrained by studying the $\nu_\mu$ disappearance of atmospheric neutrinos with the IceCube Neutrino Observatory.

\begin{figure}[!bt]\centering
\includegraphics[trim={0.5cm 0.5cm 0.5cm 0.6cm},clip,width=0.45\columnwidth]
{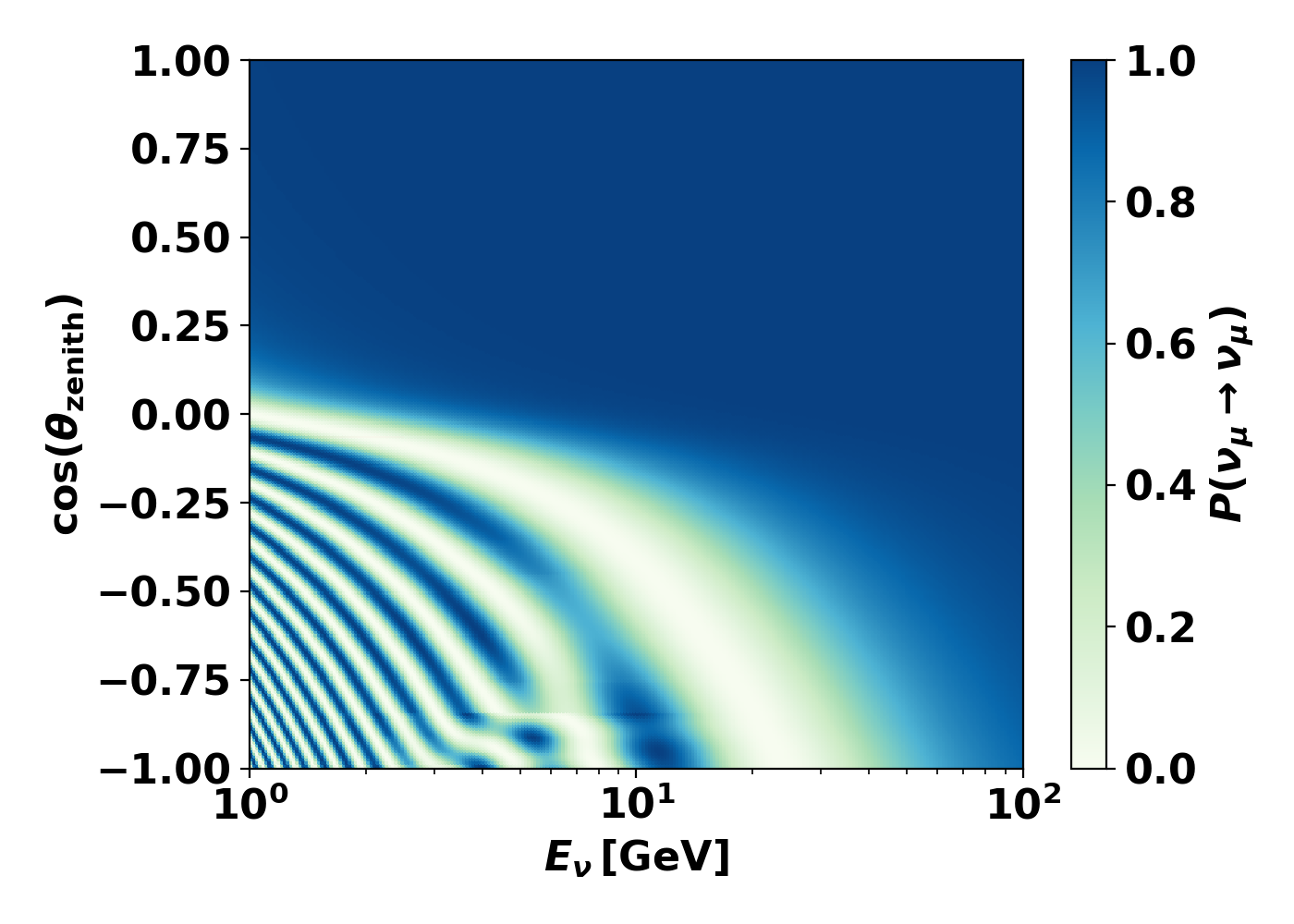}
\caption{\label{fig:osc}$\nu_\mu$ survival probability plotted against neutrino energy (E) and arrival direction (zenith angle, $\theta_{\rm{zenith}}$). The oscillation pattern (bright stripes) corresponds to atmospheric neutrinos that arrive over a broad range of baselines as they pass through the Earth.}
\end{figure}

Atmospheric neutrinos are produced in hadronic processes of cosmic ray interactions with atoms in the atmosphere. While the long-baseline experiments have fixed baselines (L), and narrow neutrino beam energy (E) ranges that are optimized for studying neutrino oscillations by looking at their L/E profiles, IceCube sees atmospheric neutrinos arrive in broad ranges of L and E, that exceed those tested by other oscillation experiments. The $\nu_\mu$ survival probability in the effective two-flavor approximation reads: \begin{equation}
P(\nu_\mu \rightarrow \nu_\mu ) \approx 1-\sin^2(2\theta_{23})\sin^2\frac{1.27\Delta m^2_{32}L}{E}.
\end{equation}
The baseline L changes with the arrival direction of atmospheric neutrinos. Figure~\ref{fig:osc} shows the expected muon neutrino survival probability in terms of neutrino energy and zenith angle. The value of $\Delta m^2_{32}$ affects the position of the oscillation ``valleys'', that are visible as bright stripes for neutrinos arriving through the Earth ($\cos\theta_{\rm zenith} \lesssim 90^\circ$) in Figure~\ref{fig:osc}.
The oscillation pattern is dominated by a neutrino sample of energy below 100 GeV arriving from below the horizon. The most outstanding oscillation minimum is at $\sim$ 30 GeV, where we expect a strong constraint on the value of $\Delta m^2_{32}$.

\section{IceCube Detector}

\begin{figure}[!tb]\centering
\includegraphics[width=0.5\columnwidth, trim=20cm 2cm 0cm 2cm]{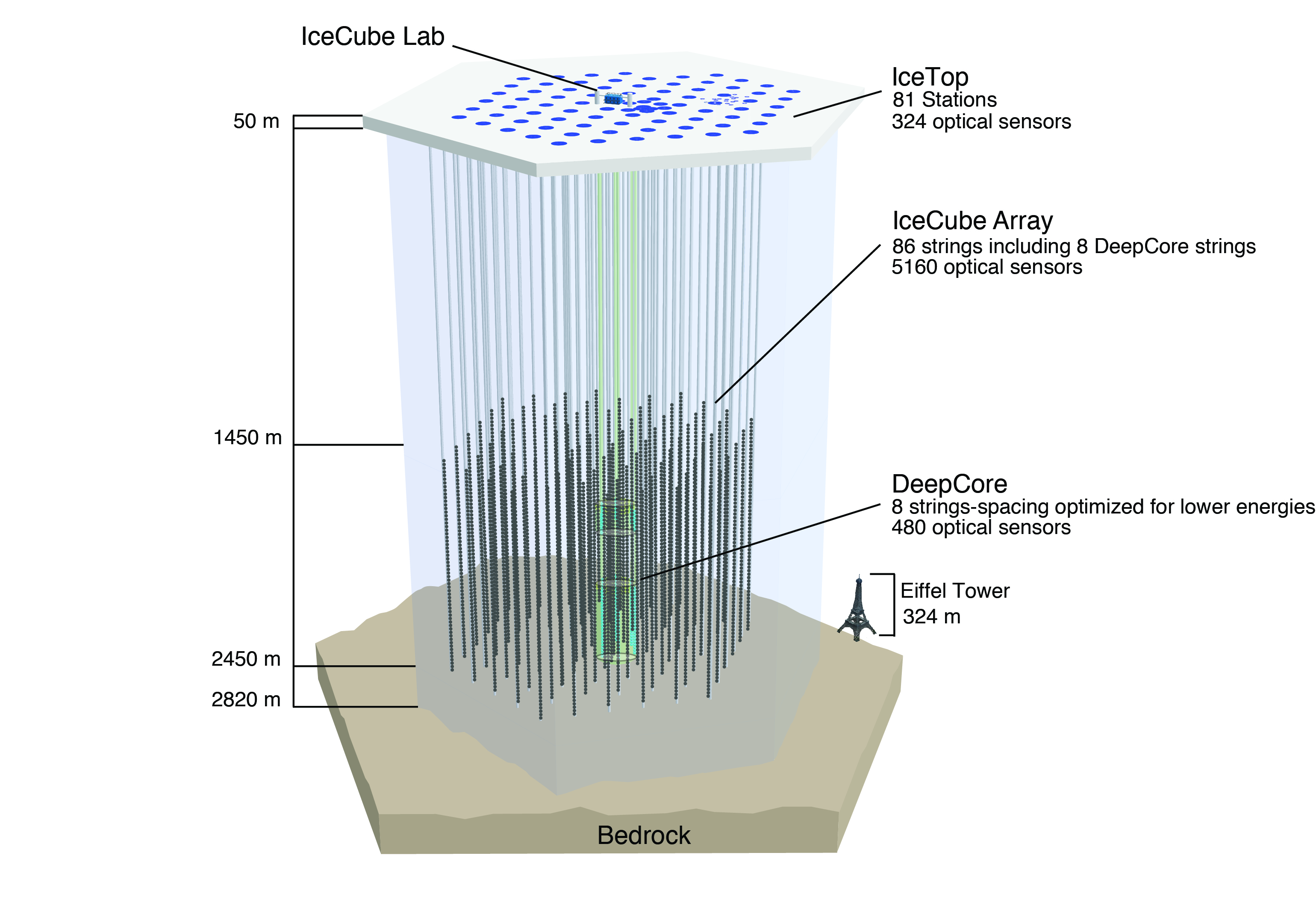}
\includegraphics[width=0.48\columnwidth, trim=1cm 1cm 5cm 10cm]{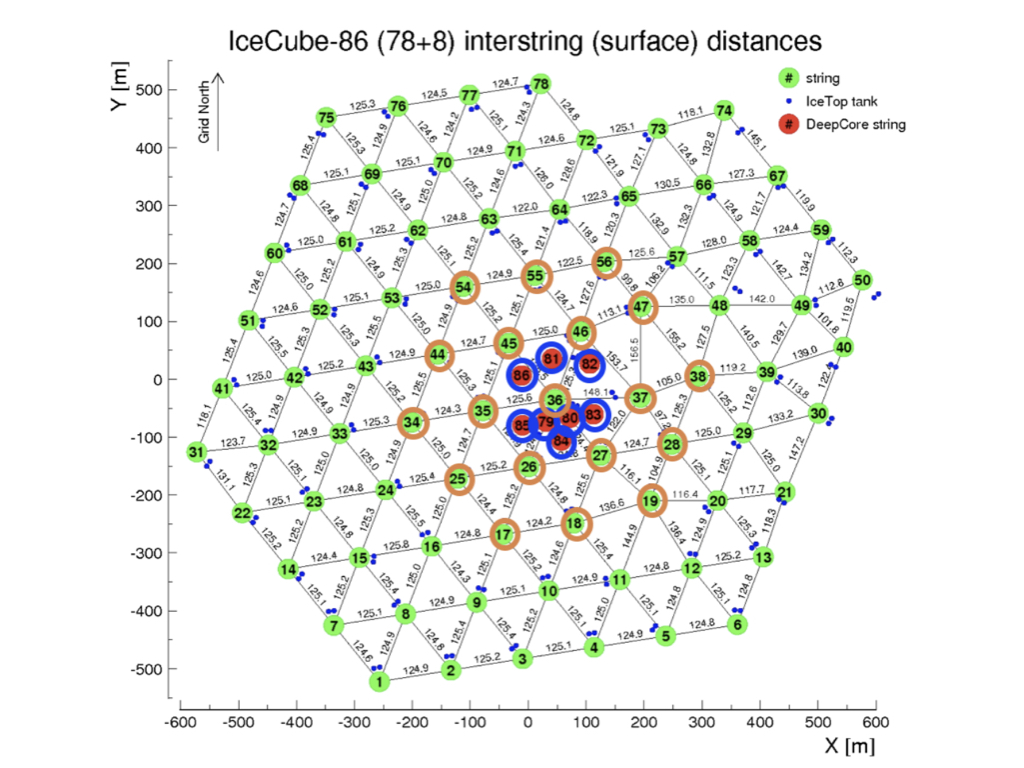}
\caption{\label{fig:detector} Sketch of the IceCube neutrino observatory showing its main in-ice array and DeepCore sub-detector (left) and its surface layout of strings (right) showing locations of the DeepCore strings (red-filled) and the surrounding IceCube main strings (orange-circled) that are used in the CNN reconstruction.}
\end{figure}

The IceCube detector (see Figure~\ref{fig:detector}) comprises 5,160 digital optical modules (DOMs) instrumenting a volume of over 1 km$^3$ of South Pole glacial ice between 1,450 m and 2,450 m deep below the surface. Each DOM is a glass pressure sphere hosting a photomultiplier tube (PMT) and the associated electronics. When neutrinos interact within the detector, the outgoing relativistic charged particles emit Cherenkov photons while propagating through the ice. The Cherenkov photons can be detected by the DOMs and converted into digitized waveforms. We extract charge and time features from the waveforms and use them as inputs to CNNs to reconstruct neutrino interactions. 78 strings composing the IceCube main detector (green dots in right panel of Figure~\ref{fig:detector}) are on a triangular grid with a horizontal spacing of about 125 m. The vertical separation of the DOMs on these strings is 17 m. There are eight strings composing the DeepCore sub-detector, located in the bottom center of the IceCube array, with a denser instrumented volume of ice ($\sim10^7$ m$^3$), enabling detection of GeV-scale neutrinos. With DeepCore, we can study neutrinos of observed energy between 5 and 100 GeV, traveling through the Earth with distances ranging up to L$\sim1.3\times10^4$ km (passing through the Earth core). 

In 2011, the IceCube detector started taking data with the fully commissioned array. This analysis uses the data taken between 2012 and 2021, excluding the data from 2011 since the detector was still stabilizing. This analysis largely inherits the simulation and calibration from the previous result with new machine-learning (ML) reconstructions employed and background-like neutrino candidates included for a better estimation of systematic uncertainties.

%\section{\label{sec:Simulation} Simulation}
%Neutrinos are generated by GENIE at version 2.12.8 \cite{genie} following a power-law flux spectrum injected surrounding the DeepCore, at a volume that is big enough to account for the events interacting outside the DeepCore but partially leaving some traces inside. The outgoing muons generated in the $\nu_\mu$ ($\bar{\nu}_\mu$) charged current interactions (CC) are propagated through matter by PROPOSAL\cite{proposal}, while $\tau$s, hadronic showers below 30 GeV, and electromagnetic (EM) showers below 100 MeV are propagated by GEANT4\cite{geant4}. For higher energy hadronic and EM showers, a GEANT4-based parameterization is applied. Cherenkov photons produced by showers and muons are propagated through the ice individually to simulate scattering and absorption\cite{clsim}. The atmospheric neutrino flux calculated for the South Pole\cite{honda} is used to reweigh the events afterward. The simulation tools and details of detector calibration methods used in this analysis are essentially identical to the previous analysis\cite{verification_paper}.

\section{\label{sec:reconstruction} Reconstruction and Event Selection}
Convolutional neural networks are employed and optimized in and near the DeepCore sub-detector, where we expect the best resolution for sub-100 GeV events, which gives us the best sensitivity for studying $\nu_\mu$ disappearance. The CNN reconstructions are used to select the final analysis sample, which helps to improve the reconstruction resolution and reduce the atmospheric muons.

\begin{figure*}[tb!]\centering
\includegraphics[trim={0.5cm 0.5cm 0.5cm .6cm},clip,width=0.5\columnwidth]
{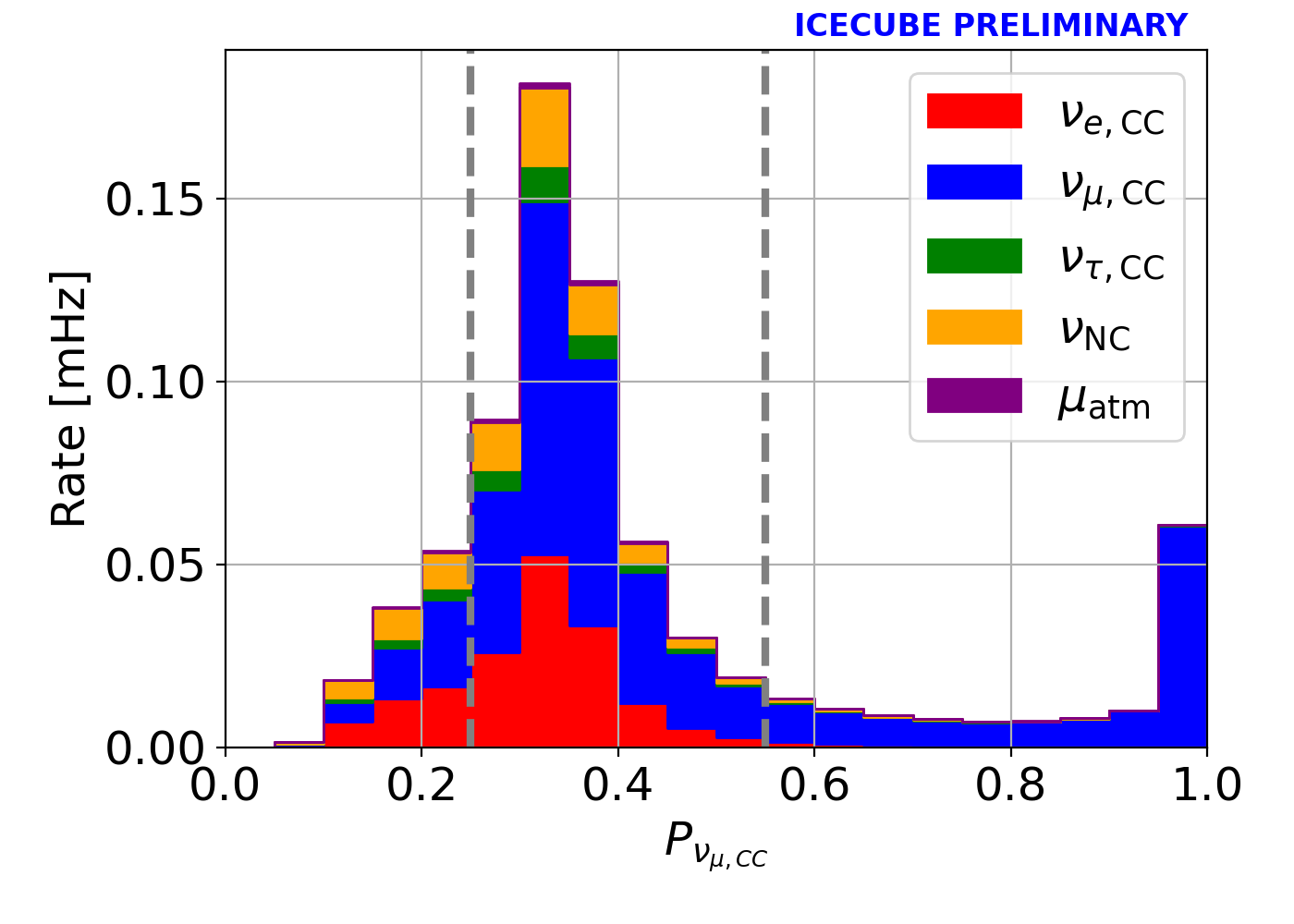}
\caption{\label{fig:pid} CNN predicted MC components. The two dashed lines indicate the boundaries between cascade- (left), mixed- (middle), and signal-like (right) events.}
\end{figure*}

We train five CNNs for neutrino energy, direction ($\theta_{\rm{zenith}}$), interaction vertex position, particle identification (PID), and atmospheric muon classification. Reconstructed energy and $\cos(\theta_{\rm{zenith}})$ are used to select neutrino candidates with reconstructed neutrino energy between 5 and 100 GeV arriving from below the horizon. The reconstructed interaction vertex helps to select neutrino candidates starting in or near DeepCore for the best reconstruction performance. The PID classifier helps to distinguish the signal events, i.e., $\nu_\mu$ charged current (CC) interactions, where the outgoing muons usually leave tracks in the detector, from the other types of neutrino interactions, whose event topologies look like scattered cascades. By recognizing the event topology, CNN can reconstruct a PID as the probability of an event being signal-like. The MC distributions of the CNN-reconstructed PID are shown in Figure~\ref{fig:pid}. 
Finally, with the help of the atmospheric muon classifier, the rate of atmospheric muon background events is well below 0.6\% of that of the entire sample. 

Each CNN is trained independently on a different sample to optimize reconstruction performance. The CNNs predicting neutrino energy and interaction vertex position are trained using signal MC events with a flat true neutrino energy distribution in 1-300 GeV with a tail extending to 500 GeV. The CNN for the zenith angle is trained on the $\nu_\mu$ CC MC events starting and ending in and near DeepCore with a flat true zenith distribution. The CNN PID identifier is trained on a sample with balanced track-like and cascade-like neutrino events. The CNN atmospheric muon classifier is trained on a balanced sample of track-like and cascade-like neutrino interactions and atmospheric muons. The CNNs use all the DOMs on the 8 DeepCore and surrounding 19 IceCube strings, as shown in Figure~\ref{fig:detector} as the orange-circled strings. Due to their different spatial densities, the DOMs on the DeepCore and IceCube strings are fed separately to the CNNs for better feature extraction and learning. The CNN reconstructions perform similarly to state-of-the-art likelihood-based reconstructions~\cite{paper_retro} while being approximately 3,000 times faster in run-time, which is a significant advantage for full MC production of large-statistic atmospheric neutrino datasets. 

\section{\label{sec:analysis}Analysis}
\begin{figure*}[tb!]
\includegraphics[trim={0.5cm 0.5cm 0.5cm 1.8cm},clip,width=\columnwidth]{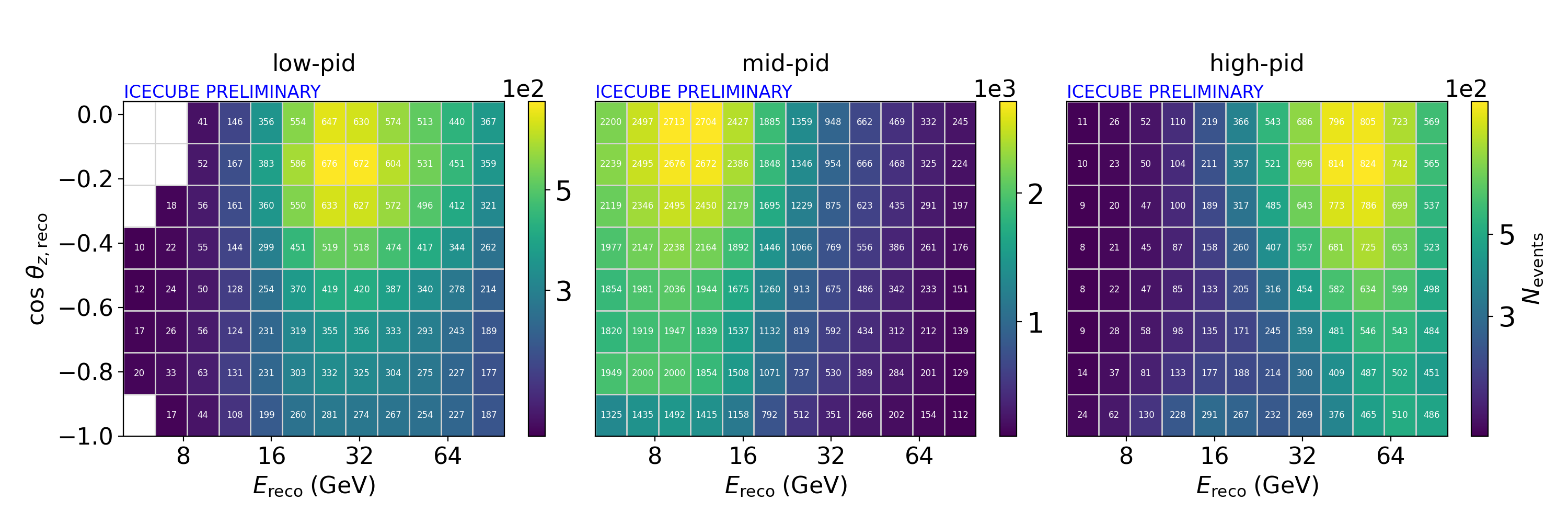}% Here is how to import EPS art
\caption{\label{fig:nominal} Selected analysis sample in bins of neutrino energy (10 logarithmic), zenith angle (8 linear), and three PID bins of cascade-, mixed-, and track-like samples from left to right. Blank bins are left out of the analysis for their low MC statistics.}
\end{figure*}

The selected sample is binned by reconstructed energy, $\cos(\theta_{\rm{zenith}})$, and PID (see Figure~\ref{fig:pid}), so that we achieve a relatively pure signal-like sample with PID$\geq0.55$ and a cascade-like sample with PID$<0.25$. The data binned in energy and $\cos(\theta_{\rm{zenith}})$ allows us to study oscillation parameters, while the three PID bins help to study systematic uncertainties.
%binning directly contribute to constraining the oscillation parameters, while PID bins help to be robust against systematic uncertainties since cascade- and track-like events are intrinsically different in detector responses, atmospheric flux origin, and cross sections. 
The binned analysis sample can be found in Figure~\ref{fig:nominal}. 

\begin{figure*}[tb!]
\centering\includegraphics[trim={0.5cm 0.5cm 1.cm 1.cm},clip, width=0.6\columnwidth]{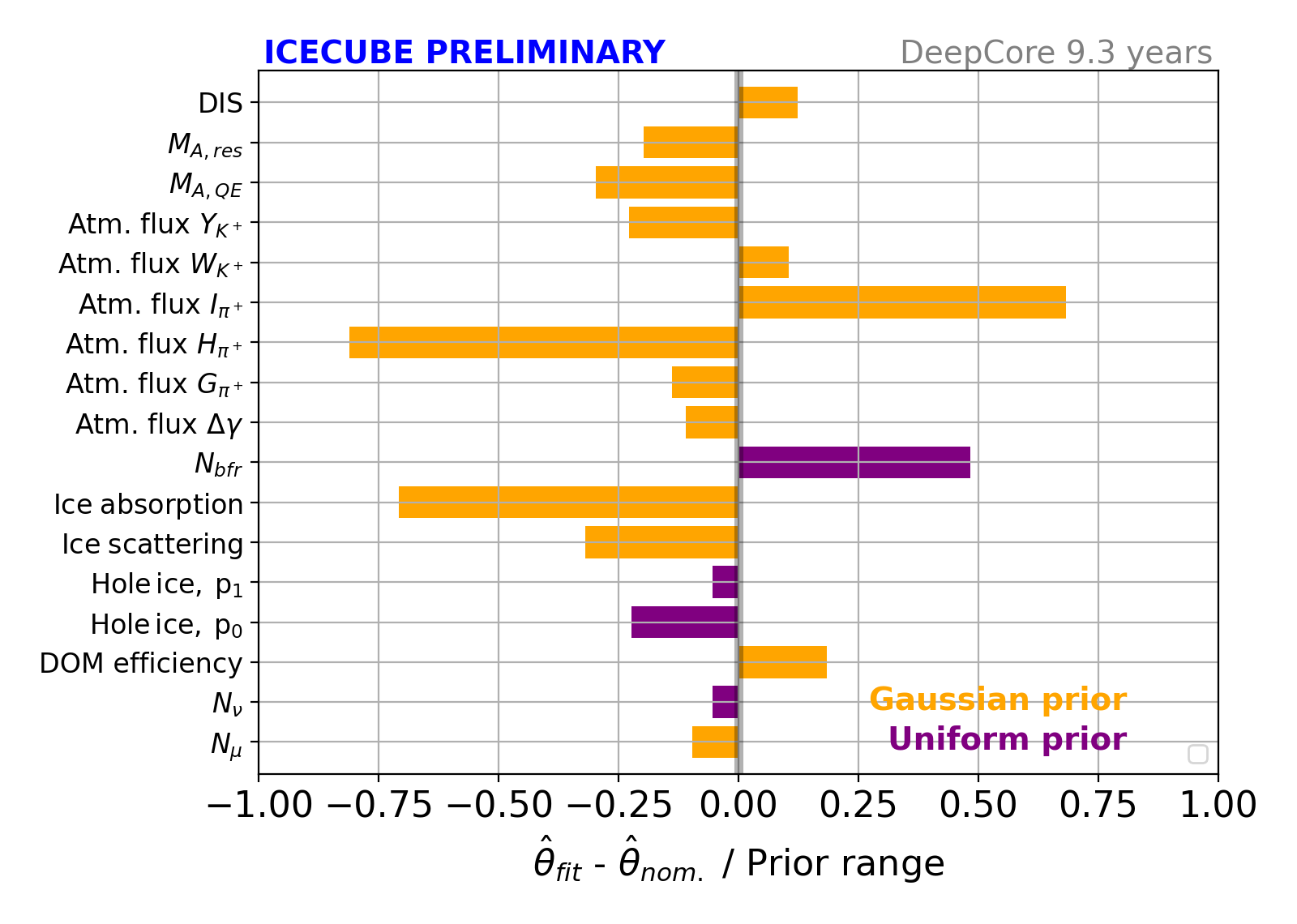}
\caption{\label{fig:sys}Fitted systematic parameters pulled from nominal values compared to ranges of priors. See the main text for the description and references of individual parameters.} 
\end{figure*}

%With different reconstructions and final sample selections present, we re-evaluated the impacts of systematic uncertainties for this analysis. 
The modeling of systematic uncertainties follows that of a previous analysis~\cite{verification_paper}. We re-evaluated their impacts on this analysis and updated the list of free systematic parameters in the fit. The fitted systematic values are compared to their nominal values and prior ranges and plotted as in Figure~\ref{fig:sys}.
Neutrino interaction cross-section uncertainties are from GENIE 2.12.8 ~\cite{paper_genie}, where the deep-inelastic scattering (DIS) parameter is interpolated between GENIE and CSMS~\cite{paper_csms}; atmospheric flux hadronic production uncertainties (``Atm.~flux Y/W/I/H/G") are employed from the work of Barr {\it{et al.}}~\cite{barr}; cosmic-ray spectral shape uncertainty is described by ``Atm.~flux $\Delta \gamma$"; ``$N_{\rm{bfr}}$" accounts for the difference introduced by the birefringent polycrystalline microstructure of the ice~\cite{bfr} compared to the ice property of nominal MC modeled by SPICE 3.2.1~\cite{verification_paper}; single-DOM light efficiency, glacial ice properties (``ice absorption" and ``ice scattering"), and refrozen ice in drilling holes (``Hole ice, p$_{0(1)}$") are largely inherited from reference~\cite{verification_paper}; and ``$N_{\nu}$" (``$N_\mu$") is responsible for the overall neutrino (muon) flux normalization. The detailed discussions of these systematic models can be found in the publication of a previous analysis~\cite{verification_paper}.

\section{\label{sec:result}Result and Conclusion}
Using the atmospheric neutrino dataset taken between 2012-2021 with a total of 150,257 events, \begin{figure*}[tb!]
\includegraphics[trim={0.5cm 0.5cm 1.55cm 1.2cm},clip, width=0.33\columnwidth]{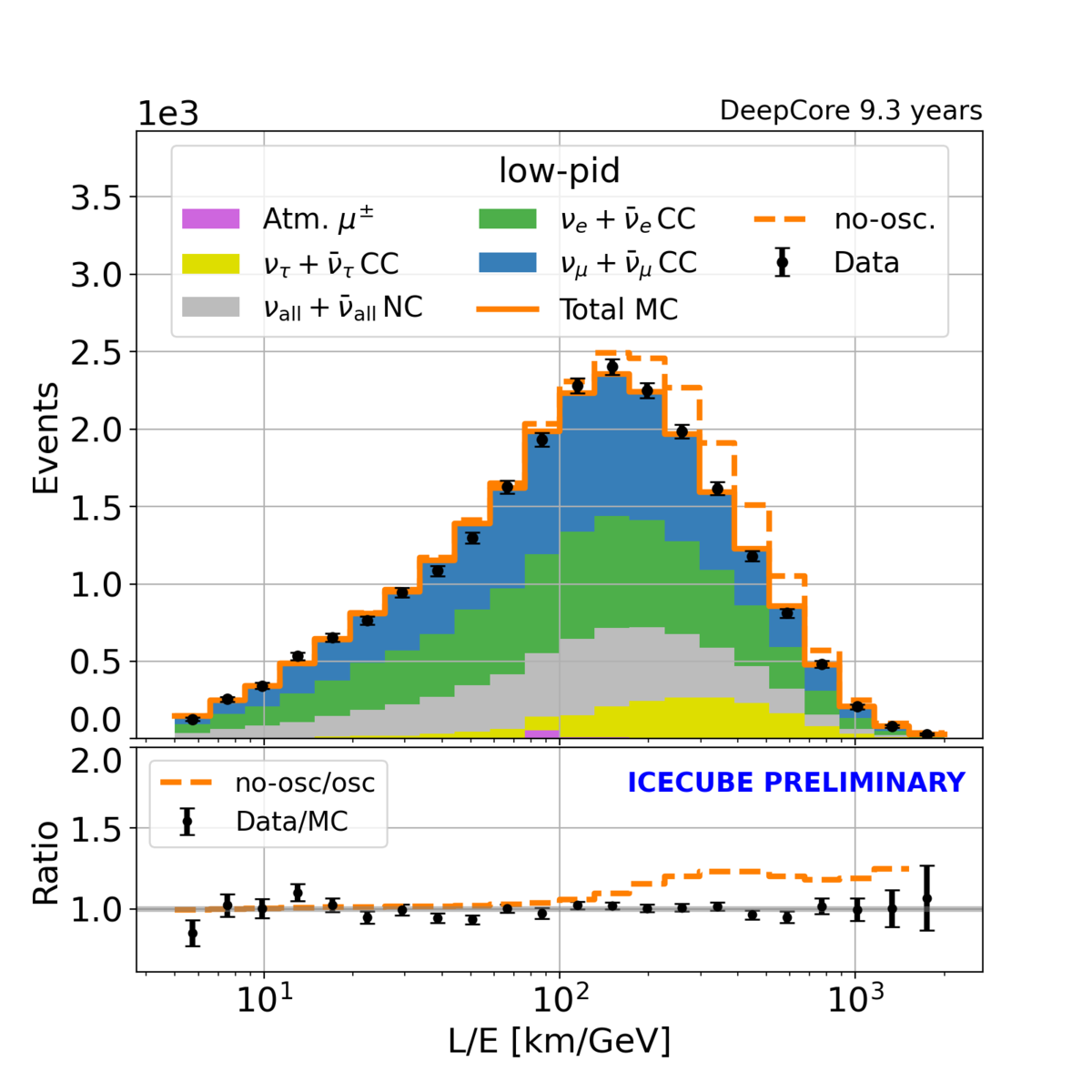}
\includegraphics[trim={0.5cm 0.5cm 1.55cm 1.2cm},clip, width=0.33\columnwidth]{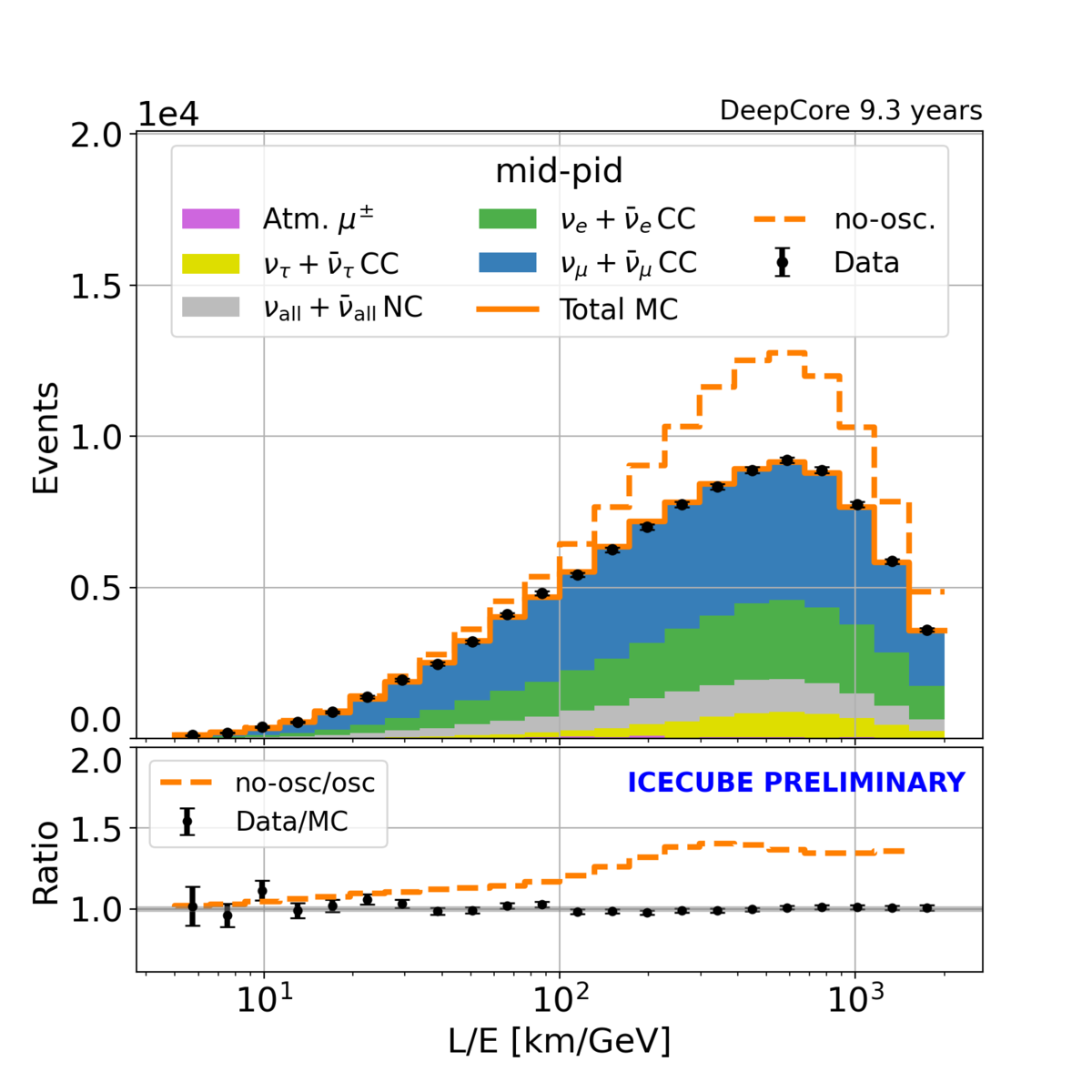}
\includegraphics[trim={0.5cm 0.5cm 1.55cm 1.2cm},clip, width=0.33\columnwidth]{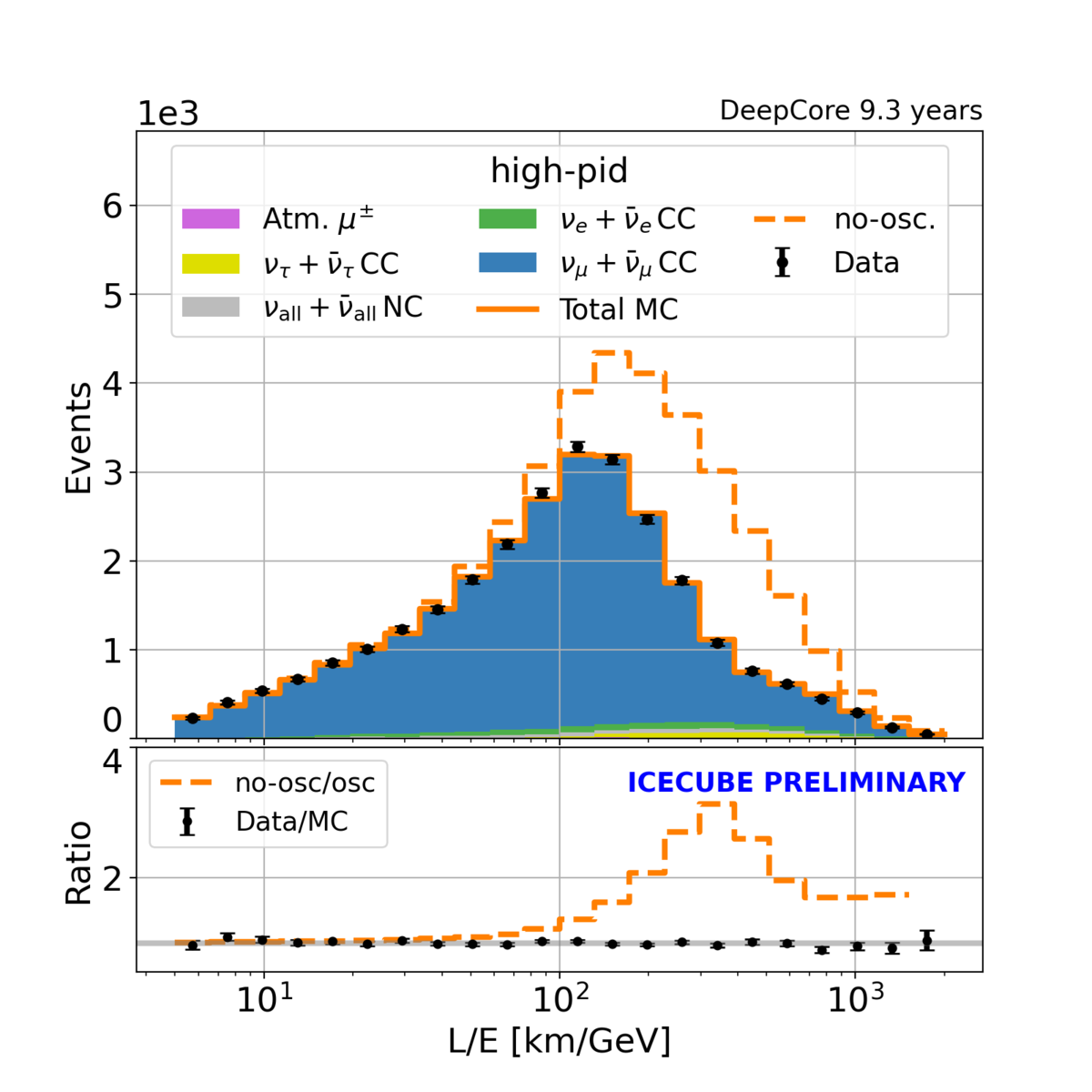}
\caption{\label{fig:lovere}Comparisons of data (black) and stacked MC L/E projections. The top panels show events in cascade- (left), mixed- (middle), and track-like (right) bins with (solid) and without (dashed) oscillations. The bottom panels show ratios of data/MC (black) and the expected ratio of MC without and with oscillations (dashed orange).} 
\end{figure*}
we achieve suitable data and MC agreement with the most outstanding disappearance signature in the signal-like (high-pid) bin, as shown in Figure~\ref{fig:lovere}.

\begin{figure*}[!tb]
\centering\includegraphics[width=0.6\columnwidth]{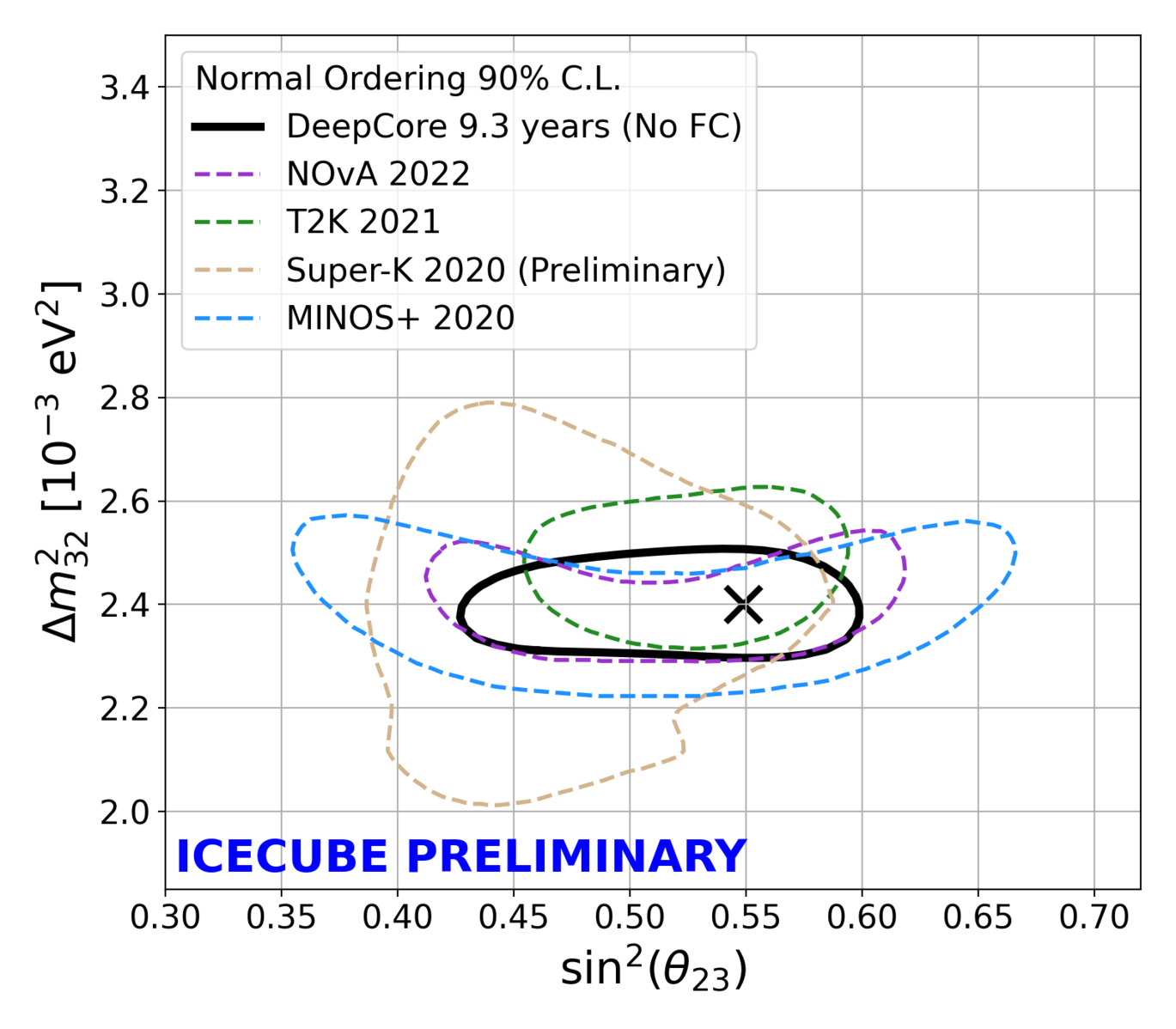}
\caption{\label{fig:contour}The 90\% C.L. contour (using Wilks' theorem and assuming normal mass ordering) and best-fit parameters (cross) of $\sin^2(\theta_{23})$ and $\Delta m^2_{32}$ compared to those of other experiments~\cite{nova_paper,t2k_paper,minos_paper,superk_paper}.}
\end{figure*}
Figure~\ref{fig:contour} shows the 90\% confidence level (C.L.) contours of $\sin^2(\theta_{23})$ and $\Delta m^2_{32}$ assuming neutrino masses are in normal ordering ($m_3 > m_2 > m_1$) of this analysis compared with those of the other experiments. 
The contour and best-fit parameters of this analysis are in good agreement with all the contours reported by the other experiments, even if we use the atmospheric neutrino dataset, which is in a relatively higher energy range and essentially different baseline range compared with the accelerator-based experiments, such as NOvA and T2K.

\begin{figure*}[!tb]
\centering\includegraphics[width=0.45\columnwidth]{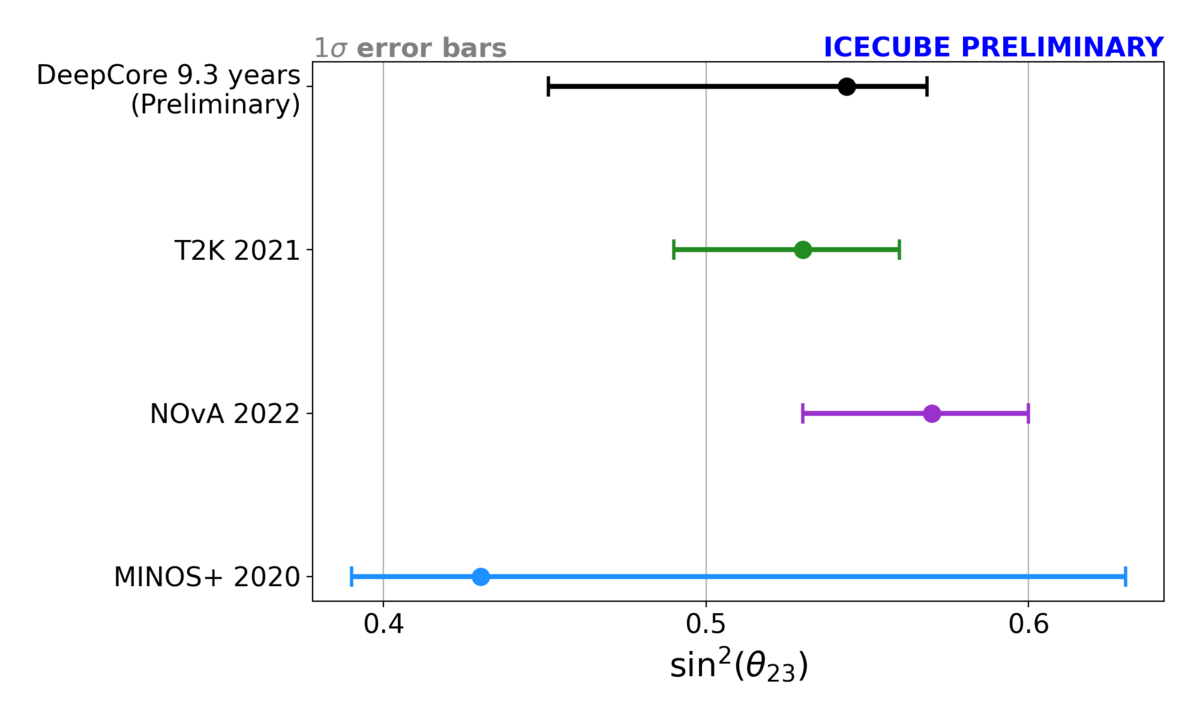}
\includegraphics[width=0.45\columnwidth]{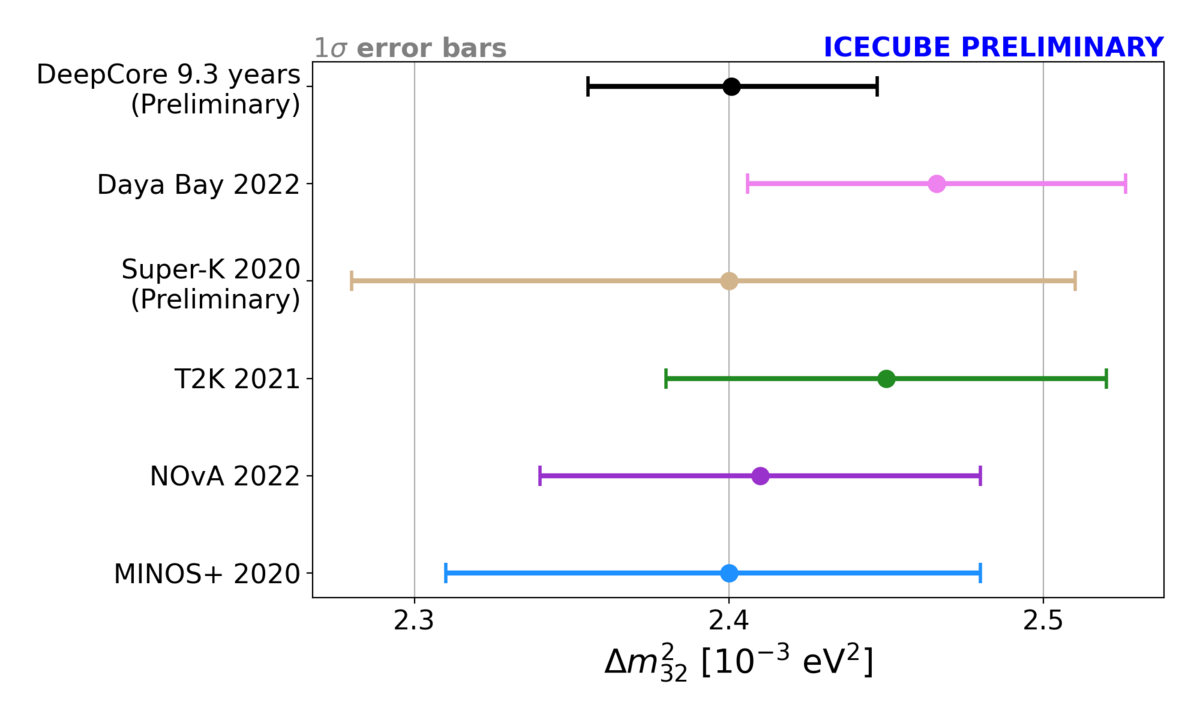}
\caption{\label{fig:errors}The 1$\sigma$ uncertainties (using Wilks' theorem and assuming normal mass ordering) of $\sin^2(\theta_{23})$ (left) and $\Delta m^2_{32}$ (right) of this analysis (black) compared with those from the other experiments~\cite{nova_paper,t2k_paper,dayabay_paper,minos_paper,superk_talk}.} 
\end{figure*}
By comparing the 1$\sigma$ uncertainties of $\sin^2(\theta_{23})$ and $\Delta m^2_{32}$ with those from the other experiments, we see good agreement among all the results on 1$\sigma$ error ranges of both parameters, as shown in Figure~\ref{fig:errors}. Additionally, our measurement of $\Delta m^2_{32}$ has a precision competitive with all previous results, which largely benefited from the most prominent oscillation minimum, as discussed earlier and shown in Figure~\ref{fig:osc}; the large-statistic atmospheric neutrino dataset; improved detector calibration and MC modeling; and the new ML reconstructions.

There are improvements in MC models underway which could lead to potential improvements in future IceCube oscillation analyses. The near-future IceCube Upgrade~\cite{upgrade} will help to improve the sensitivity by improving detector calibration. There are ongoing analyses that benefited from the CNN reconstructions and selections, such as non-standard neutrino interaction measurement and neutrino mass ordering determination. The CNN method could also be adapted and applied to data collected by the IceCube Upgrade, which will further improve the precision of measurements of neutrino oscillation parameters.

%If you more commonly use the method of square brackets in the line of text
%for citation than the superscript method,
%please note that you need  to adjust the punctuation
%so that the citation command appears after the punctuation mark.

%\subsection{Photograph}
%You may want to include a photograph of yourself below the title of your talk. A scanned photo can be  directly included using the default command\\
%\verb^\newcommand{\Photo}{\includegraphics[height=35mm]{mypicture}}^\\
%just before the \verb^\begin{document}^line. If you don't want to include your photo, just comment this line by adding the \verb^%^ sign at the beginning of the line and uncomment the next one $$\verb^%\newcommand{\Photo}{}^ by removing its \verb^%^ sign.


\begin{thebibliography}{99}

\bibitem{icecube_prd}The IceCube Collaboration, \Journal{\PRD}{99}{032007}{2019}.

\bibitem{icecube_prl}The IceCube Collaboration, \Journal{\PRL}{120}{071801}{2018}.

\bibitem{nova_paper}The NOvA Collaboration, \Journal{\PRD}{106}{032004}{2022}.

\bibitem{t2k_paper}The T2K Collaboration, \Journal{\PRD}{103}{112008}{2021}.

\bibitem{minos_paper}The MINOS+ Collaboration, \Journal{\PRL}{125}{131802}{2020}.

\bibitem{dayabay_paper}The Daya Bay Collaboration, \Journal{\PRL}{161802}{130}{2023}

\bibitem{pdg}The Particle Data Group, \Journal{Prog.\ Theor.\ Exp.\ Phys}{2022}{083C01}{2022}.
%Daya Bay collaboration, Precision measurement of reactor antineutrino oscillation at kilometer-scale baselines by Daya Bay, 	arXiv:2211.14988
\bibitem{paper_retro}The IceCube Collaboration, \Journal{Eur. Phys. J. C}{82}{807}{2022}.

\bibitem{verification_paper}The IceCube Collaboration, submitted to {\it Phys. Rev. D}.

\bibitem{paper_genie}C. Andreopoulos {\it et al}, \Journal{\NIMA}{614}{87–104}{2010}.

\bibitem{paper_csms}A. Cooper-Sarkar, P. Mertsch, and S. Sarkar, \Journal{JHEP}{08}{042}{2011}.

\bibitem{barr}G. Barr, T. Gaisser, S. Robbins, and T. Stanev, \Journal{\PRD}{74}{094009}{2006}.

\bibitem{bfr}D. Chirkin and M. Rongen on behalf of the IceCube Collaboration, \Journal{PoS}{ICRC2019}{854}{2020}.

\bibitem{superk_paper}V. Takhistov on behalf of the Super-Kamiokande Collaboration, \Journal{PoS}{ICHEP2020}{181}{2021}.

\bibitem{superk_talk} M. Gonchar on behalf of the JUNO collaboration, presentation at Nucleus-2022.

\bibitem{upgrade}A. Ishihara on behalf of the IceCube Collaboration, \Journal{PoS}{ICRC2019}{1031}{2021}.

\end{thebibliography}
\end{document}